\documentclass[prl,twocolumn,showpacs,
superscriptaddress,floatfix]{revtex4}
\usepackage{graphicx}
\usepackage{latexsym}

\begin{document}

\title{Signatures of Photon Localization}

\author{A.Z.~Genack}
\affiliation{Department of Physics, Queens College of the City
University of New York, Flushing, NY 11367, USA}

\author{A.A.~Chabanov} \affiliation{Department of Physics and Astronomy,
University of Texas, San Antonio, TX 78249, USA}

\date{10 November 2005}

\begin{abstract}
Signatures of photon localization are observed in a constellation
of transport phenomena which reflect the transition from diffusive
to localized waves. The dimensionless conductance, $g$, and the
ratio of the typical spectral width and spacing of quasimodes,
$\delta$, are key indicators of electronic and classical wave
localization when inelastic processes are absent. However, these
can no longer serve as localization parameters in absorbing
samples since the affect of absorption depends upon the length of
the trajectories of partial waves traversing the sample, which are
superposed to create the scattered field. A robust determination
of localization in the presence of absorption is found, however,
in steady-state measurements of the statistics of radiation
transmitted through random samples. This is captured in a single
parameter, the variance of the total transmission normalized to
its ensemble average value, which is equal to the degree of
intensity correlation of the transmitted wave, $\kappa$. The
intertwined effects of localization and absorption can also be
disentangled in the time domain since all waves emerging from the
sample at a fixed time delay from an exciting pulse, $t$, are
suppressed equally by absorption. As a result, the relative
weights of partial waves emerging from the sample, and hence the
statistics of intensity fluctuations and correlation, and the
suppression of propagation by weak localization are not changed by
absorption, and manifest the growing impact of weak localization
with $t$.
\end{abstract}
\pacs{42.25.Dd, 42.25.Bs, 05.40.-a}

\maketitle
\section{Overview}

Localization \cite{1,2,3,4,5} is inherently a wave interference
phenomenon \cite{6} and may therefore occur for both classical
\cite{7,8,9,10,11,12,13,14} and quantum mechanical waves. It may
be considered for all manner of excitations including
electromagnetic, sound and ultrasound, acoustic and optic phonons,
surface plasmons, polaritons, electrons and atoms. This article
will discuss signatures of electromagnetic localization in the
statistics of radiation reflected from and transmitted through
random samples.

The study of transport has been enriched by the many differences
and similarities between electronic conduction and electromagnetic
wave propagation. Differences in the types of wave and their scattering
interactions affect the types of measurements that can be carried
out, as well as the possibility and ease with which waves can be
localized. Whereas classical waves may be phase-coherent in large
static samples since wavelengths greatly exceed atomic dimensions,
electronic samples must be cooled to ultra-low temperatures in
order to suppress dephasing even in micron-sized samples. The
value of the transport mean free path, $\ell$, in which the
direction of propagation is randomized, is determined by the
specifics of the wave interaction in each material system.
However, the character of classical and quantum propagation in
mesoscopic systems, in which the wave is temporally coherent
throughout the sample \cite{5,15,16,17}, is strikingly similar on
length scales greater than both the wavelength and the mean free
path. For both quantum and classical waves, the character of
transport depends upon the closeness of the wave to the threshold
of the localization transition separating extended and localized
waves.

Ioffe and Regel proposed that propagating waves cannot exist if
scattering occurs on a length scale smaller than
$\lambda/2\pi=1/k$, in which the amplitude of the wave changes.
This suggests that propagation ceases when scattering is strong
enough that, $\ell<2\pi/\lambda$, or $k\ell<1$ \cite{6}. The
interference of waves that follow the trajectories in opposite
senses around a closed loop enhances the return of the waves and
thus suppresses transport. In weakly scattering three-dimensional
systems, the probability that a meandering wave trajectory will
loop back to a typical coherence volume,
$V_c\sim\left(\lambda/2\right)^3$, is $\sim
1/\left(k\ell\right)^2$. The impact of wave interference on
average transport is therefore small when $k\ell\gg1$. The
ensemble average of the spread of the square amplitude of the wave
is then diffusive and corresponds to the random walk of electrons
or photons. However, when $k\ell\sim1$, the probability that waves
will return to a coherence volume is substantial. Interference
then impedes the escape from this volume. This consideration leads
again to the Ioffe-Regel criterion for localization, $k\ell\sim
1$.

Measurements of $k\ell$ therefore give an indication of the
closeness to the localization threshold. Values of $k\ell$ can be
determined from measurements of the width of the peak of enhanced
backscattering, which is $\sim 1/k\ell$ \cite{18,19,20,21}. This
peak has a triangular cusp with a maximum enhancement of
retro-reflection over the diffuse background of a factor of 2 in
the diffusive limit \cite{18,19,20}, and a smaller enhancement due
to recurrent scattering events when the localization threshold is
approached \cite{21}, as seen in Fig.~1. The reflected enhancement
peak is the Fourier transform of the point spread function of the
incident beam \cite{22}.
\begin{figure}
\includegraphics [width=\columnwidth] {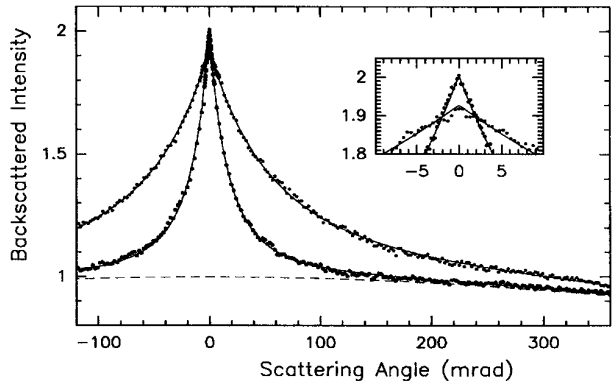}
\caption{Coherent backscattering of light measured, with two
different values for the transport mean free path $\ell$. The
typical angular width varies as $\lambda/\ell$. Narrow cone: a
sample of BaSO$_4$ powder with $\ell/\lambda=4$; broad cone:
TiO$_2$ sample with $\ell/\lambda=1$. The inset confirms the
triangular cusp predicted by diffusion theory, and also shows that
the maximum enhancement factor is lowered for the sample with
small value of $\ell/\lambda$ \cite{21}. (with kind permission of
the authors)}
\end{figure}

The closeness to localization in diffusive samples can be
ascertained from a determination of $\ell$ from measurements of
the scaling of total transmission \cite{23}. Measurements of
optical transmission through a wedge-shaped sample of random
titania particles dispersed in Polystyrene are shown in Fig.~2
\cite{23}. In the diffusive limit, the ensemble average of the
spatial intensity distribution corresponds to
the solution of the diffusion equation for the intensity. When
absorption is absent, the average energy density falls linearly
with depth inside the sample, corresponding to transmission which
falls inversely with sample thickness, $L$. When absorption is
present in a diffusive sample, the ensemble average of both the
energy density versus depth and the transmitted flux versus $L$
falls exponentially with attenuation length, $L_a=\sqrt{D\tau_a}$,
for $L>L_a$, where $D$ is the diffusion coefficient and $\tau_a$
is the absorption time \cite{23}. The variation of transmission as
$1/L$ is seen in the log-log plot in Fig.~2a for $L<L_a$, while
the exponential variation of transmission for $L>L_a$ is exhibited
in the semi-log plot in Fig.~2b. Transmission also falls
exponentially as a result of wave localization in strongly
scattering samples.
\begin{figure}
\includegraphics [width=\columnwidth] {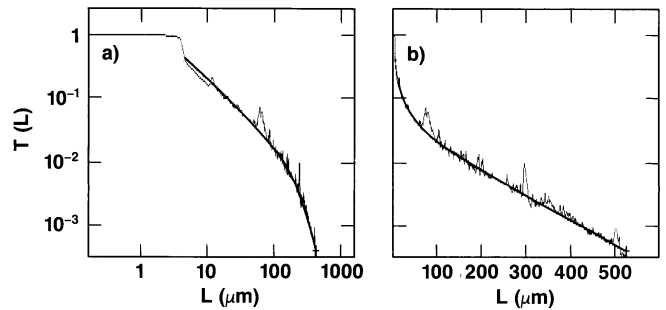}
\caption{Scale dependence of optical transmission through a
wedge-shaped sample of random titania particles dispersed in
Polystyrene. The curve gives the fit of the envelope of the data
to diffusion theory. Transmission varies as (a) $1/L$ for $L<L_a$
and as (b) $\exp\left(-L/L_a\right)$ for $L>L_a=112\pm5$ m. From
Ref.~\cite{23}.}
\end{figure}

In three-dimensional samples, the diffusion coefficient is given
by,  $D={1\over 3}v_E\ell$, where $v_E$ is the transport velocity
\cite{24}. The transport velocity is reduced when the wave is
resonant with the constituents of the sample, and exhibits dips
near Mie resonances with spherical dielectric particles
\cite{24,25,26,27}. The transport mean-free-path can be found once
$D$ and $v_E$ has been determined. The diffusion coefficient can
be determined directly from measurements of pulsed transmission
\cite{28,29,30,31,32,33} or from the correlation function of the
field or intensity with frequency shift \cite{23,24,29,34,35}. The
field correlation function is the Fourier transform of the
ensemble average of the response to a short incident pulse
\cite{30,36}. The diffusion coefficient can also be determined in
diffusive systems from the intensity correlation function with
delay time in a medium in which the constituents are in random
motion, such as in colloidal samples \cite{37,38}. The temporal
correlation function reflects the average path length distribution
in the medium, which is proportional to the average temporal
evolution of transmitted pulses.

For bounded samples, in the absence of inelastic processes,
however, any marker of localization must also depend upon the
sample geometry, in addition to $\ell$ and $\lambda$, since
reflection from the sample boundaries also returns the wave to
points within the sample. When inelastic scattering is present,
either the impact of inelastic scattering must be explicitly
accounted for or a propagation parameter must be found, which
still captures an essential aspect of localization and thereby
directly expresses the closeness of the wave to the localization
threshold.

Spectra of transmitted intensity for extended optical \cite{23}
and microwave \cite{39,40} radiation, such as the microwave
measurements shown in Fig.~3a \cite{39}, are composed of many
overlapping lines which result in spectra without isolated peaks.
The wave inside the medium, as well as each of the quasimodes, is
extended in space. Within the frequency range in which the wave is
localized, however, distinct resonances are observed in the
transmission spectrum \cite{2,41}, as seen Fig.~3b. This spectrum
of microwave transmission of localized waves was measured in a
sample of alumina spheres at low density, contained in a copper
tube, which was cooled to the temperature of liquid nitrogen to
reduce absorption losses in reflection from the tube. When the
wave resonantly excites the sample, its amplitude may be
exponentially peaked within the sample \cite{42,43,44}. The
relatively high intensity in the interior of the sample, as
compared to near the boundaries, leads to long decay times for the
excited mode and, consequently, to distinct narrow spectral lines.
Transmission via localized quasimodes is particularly high when
intensity is peaked near the middle of the sample \cite{42,43}.
Transmission is also high when two or more quasimodes occasionally
overlap spectrally leading to a multi-peaked intensity
distribution extending through the sample, known as necklace
states \cite{45,46,47}. Because these short-lived states are
spectrally broad, they dominate average transmission in 1D system.
Off resonance, the overall energy density falls exponentially
within the sample.
\begin{figure}
\includegraphics [width=\columnwidth] {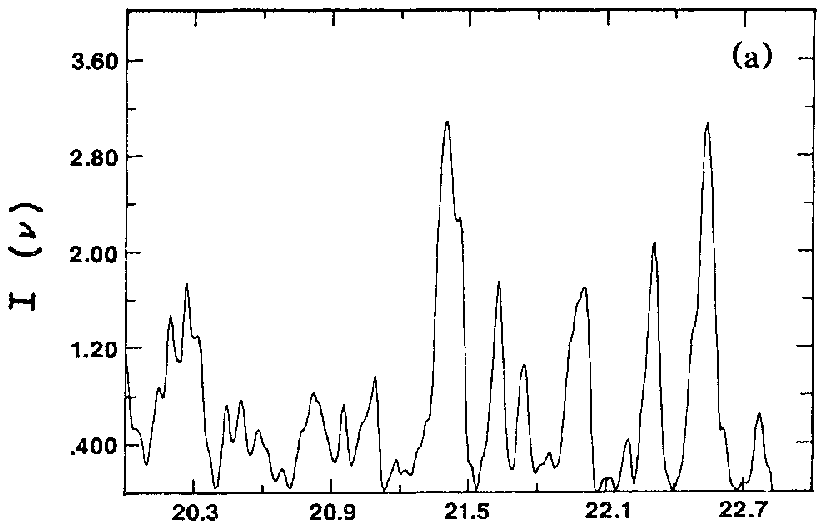}
\includegraphics [width=\columnwidth] {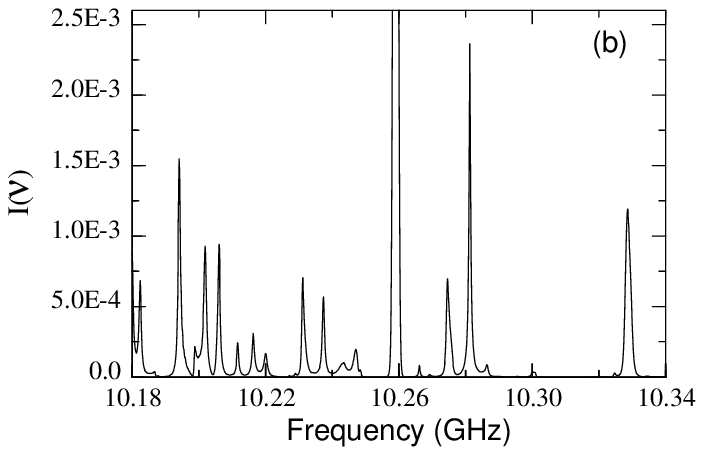}
\caption{Isolated peaks in (b), as opposed to (a), are an
indication of photon localization. The variance of relative
intensity is high in spectra with isolated peaks. From
Ref.~\cite{39}.}
\end{figure}

Localizing photons has proven to be a considerable challenge.
Photon states bound to subwavelength-scale particles do not exist
as they do for electrons bound to ionic cores. In addition,
$s$-wave scattering does not exist for photons, as it does for
electrons. The electromagnetic scattering strength due to $p$-wave
scattering vanishes in the long-wavelength, Rayleigh limit, as
$a^6/\lambda^4$, where $a$ is the particle radius. As a result, it
is not possible to enhance scattering by packing many small
scattering particles into volumes comparable to the wavelength.
Scattering for spherical particles is peaked at Mie resonances,
when the particle radius is comparable to $\lambda$. But then it
is difficult to create strong scattering with $\ell$ as short as
$\lambda$, since the maximum strength of scattering from
individual particles is achieved only in dilute systems when
scatterers are separated from one another. Because photons are
uncharged, however, a pure Anderson localization transition
between diffusing and localized waves might occur without the
complication of the Coulomb interaction. If this could be
observed, it would be possible to determine the critical exponents
of the Anderson transition \cite{48,49}. In addition, the
statistics of classical transport can be determined in large
ensembles of equivalent samples. To date, photons have only been
localized in three-dimensional systems with residual long-range
order \cite{41,50}, in one- \cite{51,52}, two- \cite{53}, and
quasi-1D \cite{41,55} samples, and in layered samples
\cite{56,57,58}.

John \cite{59} pointed out that waves can be localized in nearly
periodic samples in the frequency range of the photonic band gap
\cite{60,61} that forms in the associated periodic structure when
the index of modulation is sufficiently large. Though the density
of states vanishes in large periodic structures, the introduction
of disorder creates localized states in the spectral region of the
photonic band gap. The low density of states within the spectral
range of the photonic band gap and the narrow, spectrally isolated
lines associated with the long residence time of light on
resonance with these states lead to spatially localized modes.
Measurements of large microwave intensity fluctuations within the
low frequency band gap of a cubic wire structure have indicated
that localization occurs for intermediate filling fractions of
metallic spheres floated within the structure \cite{41,50}.

The relationship between average transmission and fluctuations in
transmission lies at the core of the scaling theory of
localization \cite{3,4,62}. In 1D, it has been hypothesized that
the relationship can be expressed via a single parameter \cite{4},
which relates fluctuations in transmission, $T$, to the ensemble
average of the logarithm of transmission, which is normally
distributed. The variance of the logarithm of the transmission in
samples of length, $L$, was conjectured to be,
$\sigma^2=\gamma/L$, where $\gamma$ is the Lyapunov exponent,
which is the inverse of the localization length,
$\gamma=1/\xi=-\ln\!T/L$, for localized waves \cite{4}. Recently
strong deviations from the single parameter scaling hypothesis
have been predicted in the tail of a band in which the density of
states is low \cite{63,64,65,66}. In addition to the localization
length, $\xi$, the scaling of transmission then depends upon a
second length scale, $\ell_S$, whenever $\xi<\ell_S$ \cite{64,67}.
This distance is the average spacing between localized states with
frequencies between the given frequency and the edge of the band
at which the density of states vanishes. The breakdown of single
parameter scaling has also been demonstrated in 2D samples
\cite{68,69,70}.

There has been considerable interest in propagation and
localization in quasi-1D samples. These samples are locally
three-dimensional, but have reflecting transverse boundaries with
length, $L$, considerably exceeding the sample's transverse
dimensions. Since all modes are mixed in such systems, the
statistics of propagation are the same at any point on the output
surface. Random matrix theory \cite{71,72} is well suited to such
systems and has provided nonperturbative results that supplement 
the findings of diagrammatic perturbation theory in the weak
scattering limit \cite{5,73}. At the same time, measurements of
field, intensity, and total transmission in the frequency and time
domains can be performed on random ensembles of quasi-1D samples.
Because of reflection at the transverse boundaries, the wave is
inhibited from straying far from a point in the transverse
directions. The number of returns to a typical coherence volume,
$V_c\sim\left(\lambda/2\right)^3$, increases superlinearly, as $L$
increases. As a result, the wave can always be localized as the
sample length is increased while the cross sectional area remains
fixed. This makes it possible to obtain a comprehensive
statistical picture of propagation and localization in samples
with well-defined statistical parameters in diffusive and
localized samples \cite{74}.

Key properties that lend themselves to a universal description are
the average values of transmission and conductance and their
scaling, and the correlation and variance of local and spatially
averaged flux or current. A variety of parameters capture the
changing character of diverse aspects of wave propagation in the
localization transition. The relationship between these phenomena
and the parameters that characterize them are explicit in the
absence of inelastic processes, since transport may be described
in terms of a single parameter near the center of a band.
Indications of a breakdown of single parameter scaling in regions
in which the density of states is low have not been found in
measurements in locally 3D samples in the slab or quasi-1D
geometries. When absorption is present, however, scaling cannot be
described in terms of a single parameter, and the resultant
exponential reduction in transmission is not related to increasing
localization of the wave. Though the conductance can then no
longer serve as a localization parameter, measures of correlation
and fluctuations of transmitted flux may still be expressed
through a single parameter which is a reliable indicator of the
nature of wave transport.

We will find that a clearer picture of wave propagation can be
achieved once the effects of absorption and weak localization are
disentangled. This can be done by transforming from the frequency
to the time domain, since absorption has the same impact on all
paths at a given delay from an exciting pulse. Since the
distribution of paths is not changed by absorption, relative
fluctuations are unchanged at a fixed delay time $t$ from an
exciting pulse. The temporal evolution of the wave can also be
explored by studying its Fourier transform, the field correlation
function with frequency shift.

In the absence of inelastic processes, the suppression of
transport due to the enhanced return of the wave to a coherence
volume may be described in terms of the average value of the
dimensionless conductance $g$, which is the scaling parameter
\cite{3,4} of the Anderson localization transition,
$g=G/\left(e^2h\right)$. In a particular sample realization, the
dimensionless conductance equals the transmittance $T$, which is
the sum of transmission coefficients over all $N$ input transverse
modes, $a$, and outgoing modes, $b$, $T=\Sigma_{ab}T_{ab}$
\cite{75,76,77}. The conductance is naturally measured in
electronic experiments since the relative phase of the incident
transverse current modes fluctuates randomly over the measurement
time to yield the incoherent sum of transmission coefficients. In
contrast, studies of classical waves are often carried out with
coherent sources, so that it is natural to measure either the
field or flux transmission coefficient for a single incident mode
to a single outgoing mode or the total transmission which is the
transmission coefficient for the net transmitted flux for a single
incident transverse mode.

In quasi-1D samples, the number of transverse modes in which two
transverse wave polarizations are distinguished is $N=2\pi
A/\lambda^2$, where $A$ is the sample cross section. In the
diffusive limit, $g\gg 1$, the dimensionless conductance can be
approximated by $g=N\ell/L$. This is equivalent to Ohm's law for
diffusing particles, since
$g=\left(2\pi\ell/\lambda^2\right)\left(A/L\right)$. To order
$1/g$, the conductance is reduced from the expression above by a
factor of $\left(1-1/3g\right)$ \cite{78}. Scaling of average
transport is given in terms of the single parameter, $g$, and the
localization threshold is reached at $g\approx 1$ \cite{4}, at
which point $g$ is substantially reduced by renormalization due to
weak localization. Since the value of $g$ falls at least as fast
as it does for diffusing particles, it is possible to localize
waves in static systems by choosing sample lengths greater than
the localization length, $\xi=N\ell$. This follows the analysis by
Thouless that electrons can be localized in sufficiently long
wires when the temperature is low enough that dephasing can be
neglected \cite{2}.

The relationship of $g$ to localization may be seen from its
connection to the Thouless number, $\delta$ \cite{2}, which may be
expressed as the ratio of the typical width and spacing of
resonances of a random ensemble of samples,
$\delta=\delta\nu/\Delta\nu$. The level width is the average
leakage rate of energy from the electromagnetic quasimodes of the
sample and is essentially the inverse of the width of the
distribution of photon transit times through the sample when
inelastic processes are absent. This corresponds to the width of
the field correlation function with frequency shift \cite{79} and
is given by $\delta\approx D/L^2$ \cite{2}. The level spacing is
the inverse of the density of states of the sample as a whole, and
may be written as $\Delta\nu=1/n(\nu)AL$, where, $n(\nu)$ is the
number of states per unit volume per unit frequency, and $AL$ is
the sample volume. The Thouless number is a measure of
localization since transport is then suppressed by the poor
overlap of quasimodes in different sectors of a sample when,
$\delta\nu<\Delta\nu$, and hence, $\delta<1$. When $\delta>1$, on
the other hand, quasimodes overlap spectrally and transport is
relatively uninhibited. Using the Einstein relation for the
conductivity, $\sigma=\left(e^2/h\right)Dn(\nu)$, the conductance
can be written as  $G=\sigma A/L=\left(e^2/h\right)Dn(\nu)A/L$. We
can then write,
$\delta=\delta\nu/\Delta\nu=\left(D/L^2\right)/\left(
1/n(\nu)AL\right)=Dn(\nu)A/L=G/\left(e^2/h\right)=g$ \cite{2}.

The degree of correlation can be identified with $1/\delta$,
since, for monochromatic excitation, the number of quasimodes
which are substantially excited is approximately the number of
modes within the mode linewidth, which is
$\delta\nu/\Delta\nu=\delta$. Since the state of the wave in the
sample is approximately specified by the amplitudes of these
$\delta$ modes, it may be specified approximately by $\delta$
parameters. As a result, the degree of correlation of intensity is
approximately, $1/\delta$, or equivalently, $1/g$.

In the diffusive limit, $g\gg 1$, far from the localization
threshold, the change in average conductance and the degree of
long-range correlation in the flux due to wave interference are
both of order $1/g$. Nonetheless, fluctuations in spatially
averaged flux are dramatically enhanced over the values that would
be obtained if long-range correlation were absent and the flux in
each of the $N$ transmission modes could therefore be assumed to
be statistically independent. If this were the case, the variance
of total transmission normalized to its ensemble average value,
$s_a=T_a/\langle T_a\rangle$, would be given by var$(s_a)=1/N$,
while the variance of the conductance normalized to its ensemble
average, $s=T/g$, would be given by var$(s)=1/N^2$, when
absorption is absent. Instead, the variance of normalized total
transmission over an ensemble is enhanced by a factor of $L/\ell$,
to give var$(s_a)\approx(1/N)(L/\ell)\approx 1/g$
\cite{35,80,81,82,83,84}. The variance of the normalized
dimensionless conductance is enhanced by a factor of $(L/\ell)^2$,
to give
var$(s)\approx\left(1/N^2\right)\left(L/\ell\right)^2\approx
1/g^2$ \cite{16,17,81}. This leads to a fixed value for the
variance of the dimensionless conductance,
var$(T)=g^2\,$var$(s)\approx 1$, where, $T=s\times g$.
Calculations, in accord with measurements, give var$(T)=2/15$,
independent of the size or local scattering strength of the sample
\cite{16,17,81}. This is termed universal conductance
fluctuations.

The neglect of absorption in calculations of propagation makes
it possible to highlight specific properties related to weak
localization and mesoscopic fluctuations. However, this is only
rarely justified in practice. Indeed, the attempt to deal with the
reality of absorption has sharpened our understanding of the
nature of localization and has provided clear signatures of
localization.

\section{Ensemble average transmission and reflection}

Early discussions of photon localization were stimulated by the
analogy between electron and photon transport. The impact of
localization on the scaling of transmission was represented as a
renormalization of local scattering parameters such as the mean
free path or diffusion coefficient \cite{Wolfle}. It was conjectured that the
renormalization of optical transport parameters would be cut off
by absorption just as electron renormalization is cut off by
dephasing \cite{10}. Microwave measurements of the variation of
transmission with thickness in strongly scattering mixtures of
metal and dielectric spheres showed departures from diffusive
behavior that were interpreted in terms of a scale dependent
diffusion coefficient whose value saturated in samples longer than
the attenuation length due to absorption \cite{85,86}.

On length scales shorter than the absorption length, the effective
diffusion coefficient at the mobility edge is predicted to scale
as $1/L$ and the transmission as $1/L^2$, while the transmission
is expected to fall exponentially for localized waves
\cite{10,11}. Such scaling was observed in microwave measurements
\cite{85,86}, but absorption also leads to a more rapid fall-off
of transmission in diffusive media \cite{23}. Subsequently,
similar scaling was observed in optical measurements in very
weakly absorbing and strongly scattering samples of GaAs powders
together with a broadening of the coherent backscattering peak
\cite{87}. But the role of absorption might still be significant
enough to affect these measurements in view of the lengthened
dwell time for waves in strongly scattering samples \cite{88}. The
closeness to the localization threshold in such systems can be
ascertained, however, in a variety of statistical properties of
the scattered wave, which will be discussed below.

The interaction of absorption and localization was probed in
computer simulations of the temporal variation of the spatial
distribution of acoustic energy in an absorbing 2D random sample
that is isolated from its surroundings. The vibrational amplitude
showed an exponential fall-off with displacement from the single
line which was initially vibrating, that persisted in time
\cite{89}. This demonstrated that though net energy transport is
reduced by absorption, the wave is localized and does not spread
throughout the sample. This is understandable since partial waves
following a loop in two opposite senses are in phase, when these
return to the initial scattering center. Thus the impact of weak
localization is undiminished for the surviving partial waves.

The time dependence of flux within a medium can be treated
separately from the loss of energy due to dispersed absorption.
The statistical impact of absorption at any time can be accounted
for by simply multiplying all waves by the factor
$\exp\left(-t/\tau_a\right)$. This was demonstrated in
measurements of field spectra of microwave radiation transmitted
through a random ensemble of absorbing dielectric quasi-1D samples
\cite{41}. The measured probability distribution of intensity was
shown to be in accord with a single-parameter distribution, which
is the measured variance of the intensity, as will be discussed
below. It was further shown that the impact of absorption could be
undone in a statistical sense by simply restoring the magnitude of
the amplitude of all partial waves emerging from the sample to the
value these would have had if absorption were absent without
altering the wave. This is done by transforming field spectra into
the time domain by multiplying each spectrum by a Gaussian
spectrum associated with the Gaussian temporal profile of the
incident field amplitude. This product spectrum is then Fourier
transformed to give the temporal response to an incident pulse
with the prescribed bandwidth. The field in the time domain is
then multiplied by $\exp\left(t/2\tau_a\right)$ and transformed
back to the frequency domain. The field transmission coefficient
is then obtained by dividing by the Gaussian field spectrum. The
complex square of this spectrum is then computed to give the
intensity spectrum from which the probability distribution of
transmitted intensity is determined. The resultant probability
distribution of steady-state intensity was found to be in
excellent agreement with the prediction for samples without
absorption \cite{41,73,90,91,92}. The distribution is then given
in terms of the single parameter, $g$. This demonstrates that 
statistically the impact of absorption at any time is simply to 
reduce the wave amplitude. Absorption thus has a straightforward
affect on transport in the time domain, and does not affect weak
localization or the associated renormalization of scattering
parameters at a specific delay time $t$. This is confirmed by our
finding that samples with identical scattering strengths but
different values of absorption have the same degree of correlation
at a given delay time. Indeed, correlation is not suppressed by
the increasing role of absorption for longer time delays, but is
rather enhanced with increasing time delay. The reduction of
steady-state intensity correlation in the presence of absorption
is thus seen to be a result of the decreasing weight of longer
paths with increasing absorption. Finally, this analysis shows
that the diffusion coefficient is independent of the value of the
absorption coefficient.

The role of weak localization increases with time since the
probability that a wave trajectory will cross itself increases. We
expect therefore that transport will be increasingly suppressed
with time even in nominally diffusive samples with $g>1$. An
alternate picture of dynamic propagation would be that the initial
pulse excites a distribution of quasimodes in the sample. The
decay rate of transmission would then decrease in time since the
amplitude of longer-lived quasimodes, which are more strongly
peaked near the center of the sample, would increase relative to
that of shorter-lived resonances. Averaged over an ensemble of
random sample realizations, the distribution of leakage rates for
quasimodes would be a continuum and would give rise to a
continuous reduction of the leakage rate from the sample. Both of
these pictures of pulse dynamics are at variance with the particle
diffusion picture that gives the ensemble average of intensity as
a discrete sum over diffusion modes. The decay rates of the
diffusion modes are given by
$1/\tau_n=n^2\pi^2D/\left(L+2z_0\right)^2$, where $n$ is a
positive integer and $z_0$ is the boundary extrapolation length
\cite{93,94,95}. After a time $\tau_1$, the intensity distribution
would settle into the lowest diffusion mode and would decay at the
rate, $1/\tau_1=\pi^2D/\left(L+2z_0\right)^2$.

Non-exponential decay has been observed in acoustic scattering in
reverberant rooms \cite{96,97} and solid blocks \cite{98,99} as
well as in microwave scattering in cavities whose underlying ray
dynamics is chaotic \cite{100,101}. The decay rate of electronic
conductance has been predicted to fall as a result of the
increasing weight of long-lived, narrow linewidth states
\cite{102,103}. The leading correction to the diffusion prediction
for the electron survival probability $P_S(t)$ was calculated by
Mirlin \cite{104} using the supersymmetry approach \cite{105} to
be,
$-\ln\!P_S(t)=\left(t/\tau_1\right)\left(1-t/2\pi^2g\tau_1\right)$.

In pulsed microwave measurements, in quasi-1D random dielectric
media composed of random mixtures of low-density alumina spheres,
we find a breakdown of the diffusion model in the non-exponential
decay of transmission \cite{33}. The decay rate for the
transmitted pulses shown in Fig.~4a for different sample ensembles
is seen in Fig.~4b to decreases in time at a nearly constant rate.
A linear fall of the decay rate would be associated with a
Gaussian distribution of rates of decay for quasimodes. We find a
slightly more rapid decrease of the decay rate which is associated
with a slower than Gaussian fall-off of the decay time
distribution. The temporal variation of transmission can also be
described in terms of the growing the impact of weak localization
on the dynamic behavior of waves, which can be expressed via the
renormalization of a time-dependent diffusion constant or mean
free path \cite{33,78,106,107,108}.
\begin{figure}
\includegraphics [width=\columnwidth] {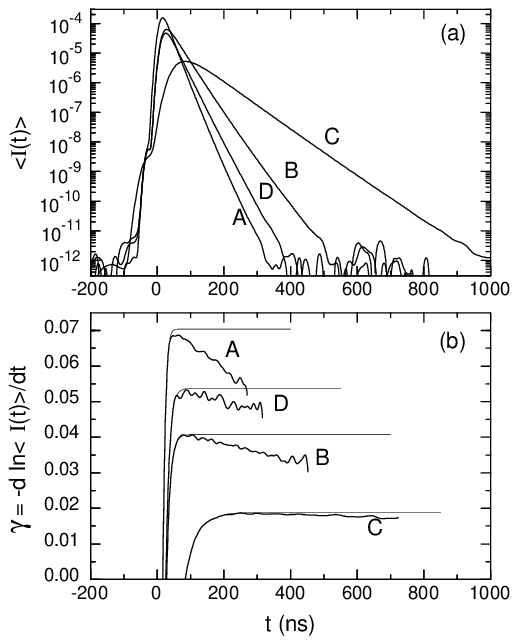}
\caption{(a) Average pulsed transmitted intensity in alumina
samples of $L=61$ (A), 90 (B and D), and 183 cm (C); Sample D is
the same as Sample B except for the increased absorption by
titanium foil inserted along the length of the sample tube. (b)
Temporal derivative of the logarithm of the intensity gives the
rate $\gamma$ of the intensity decay due to both leakage out of the
sample and absorption. The dashed curves are the decay rates given
by the diffusion model. From Ref.~\cite{33}.}
\end{figure}

Transmission for localized waves is via resonances with isolated
localized states or via spectrally and spatially overlapping
states \cite{45,46,47,58}. Steady-state measurements of transmission 
through localized states have been made in stacks of glass
slides \cite{57} and in alternating solid and porous silica layers
\cite{58}. Both overlapping and isolated spectral lines are
observed. The presence of overlapping lines in the transmission
spectrum of a layered material may have a number of causes. When a
plane wave enters a layered sample in which isotropic layers are
homogeneous, transmission can be described by a 1D model. When
$\xi<L$, then, $\delta<1$, and spectral lines are generally
distinct. However, occasional overlap may occur. This has a strong
impact on average transmission since the spatial distribution
within the sample is quasi-extended and the transmission is
consequently high \cite{46,47}. In samples with transverse
inhomogeneity, it is necessary to go beyond a 1D model. Spectral
overlap is then produced by broadening associated with the widths
of the distributions of the wave vectors in the incident beam and
of the structure factor of the layered medium with transverse
inhomogeneity. These additional factors give contributions to the
$k$-vector distributions which are separately proportional to the
inverses of the width of the incident beam and the transverse
coherence length of the sample. The overlap of modes is clearly
seen in the complex temporal evolution of the pulse in layered
porous silica samples \cite{58}. Microwave measurements on a
slotted single-mode waveguides packed with random slabs show the
impact of overlapping quasimodes within the sample \cite{52}.

Lasing is facilitated in localized modes because the excitation is
able to penetrate deep into the sample with greatly amplified
intensity when the pump laser is resonant with an exponentially
localized mode \cite{57}. Light is subsequently emitted into
long-lived modes overlapping the region of excitation. This
enhances the opportunity for stimulated emission and therefore
lowers the laser threshold. In contrast, in diffusive media \cite{110},
excitation is restricted to the surface region by multiple
scattering and the lasing threshold is not substantially reduced
below the threshold for amplified stimulated emission in
homogeneous media \cite{109}. Localized modes and low threshold
lasing were observed in stacks of glass slides with interspersed
dye solution or dye-doped plastic sheets \cite{57}. Sharp-line
random lasing spectra are also observed in diffusive media
\cite{111, 112}, in which lasing is initiated in a single or small
number of long-lived modes \cite{54,113,114,115}. Lasing also
occurs in localized states in nearly periodic anisotropic 1D
sample. This is observed in disordered cholesteric liquid
crystals. In the ordered structures, the director pointed along
the average molecular axis in each layer rotates with well-defined
pitch. A distinctive progression of band edge modes is then
observed. But in liquid crystals in which periodicity has been disrupted 
to a lesser or greater extent, localized states form at the band edge. 
Lasing then occurs in these states \cite{116,117}. Lasing is also observed 
when a localized state is introduced into the band gap by a
discontinuous jump in the angle of the director \cite{118,119}.

\section{Steady-state statistics}

In the absence of absorption, there is an explicit connection
between average transport and mesoscopic fluctuations and
correlation in quasi-1D samples since these can all be expressed
in terms of a single parameter, $g$. Since this direct link
between the averaged transmission and the statistics of
transmission is lost in absorbing samples, it is worthwhile to
consider the statistics of relative fluctuations. We consider the
probability distributions of transmitted intensity and total
transmission normalized by their ensemble averages,
$s_{ab}=T_{ab}/\langle T_{ab}\rangle$ and $s_{a}=T_{a}/\langle
T_{a}\rangle$, respectively. The relationship between the
probability distributions of intensity and total transmission was
found by Kogan and Kaveh from random matrix theory \cite{90},
\begin{equation}
P(s_{ab})=\int_0^{\infty}{ds_a\over
s_a}P(s_a)\exp\left(-s_{ab}/s_a\right). \label{}
\end{equation}
This is equivalent to the relationship between the moments of the
distributions of $s_a$ and $s_{ab}$,
\begin{equation}
\langle s^n_{ab}\rangle=n!\langle s^n_a\rangle, \label{}
\end{equation}
for the $n$-th moment \cite{90}. From the relationship between the
second moments, obtained from Eq.~(2) for $n=2$, we find,
var$(s_{ab})=2$var$(s_a)+1$. In the limit of weak scattering and
vanishing absorption, the distribution of total transmission is
found from diagrammatic and from random matrix theory to be
\cite{90,91},
\begin{equation}
P(s_a)=\int_{-i\infty}^{i\infty}{dp\over2\pi
i}\exp\left[ps_a-\Phi(p)\right], \label{}
\end{equation}
where,
\begin{equation}
\Phi(p)=g\ln^2\left(\sqrt{1+p/g}+\sqrt{p/g}\right). \label{}
\end{equation}
This leads to a connection between fluctuations and average
transmission, var$(s_a)=2/3g$. Since the probability distributions
of intensity and total transmission are given in terms of the
single parameter, $g$, in this limit, these distributions may
equally be expressed as functions only of var$(s_a)$.

The distributions of $s_a$ and $s_{ab}$ in three quasi-1D samples
composed of randomly positioned Polystyrene spheres with different
diameters and lengths in the frequency range 16.8-17.8 GHz, in
which the variation in the mean free path is small, are shown in
Fig.~5 \cite{92,120}. The distributions broaden, and the deviation
from a Gaussian become more pronounced, as either the sample
length increases or the cross-sectional area decreases. The
results shown in Fig. 5 are for samples, $a$, $b$, and $c$, which
would have values of the dimensionless conductance in the absence
of localization corrections and absorption, $g=N\ell/L$, of 15.0,
9.0, and 2.25, respectively \cite{92}. The straight line in
Fig.~5b is a plot of $\exp(-s_{ab})$, which is the prediction of
Rayleigh for the intensity distribution for one polarization
component of the field in a sample in which the number of
transverse modes, $N$, diverges and in which long-range
correlation vanishes. The distribution of total transmission in
the diffusive limit is Gaussian with variance $2/3g$. This is in
contrast to an estimated value of $1/N$, if long-range
correlation were neglected.
\begin{figure}
\includegraphics [width=\columnwidth] {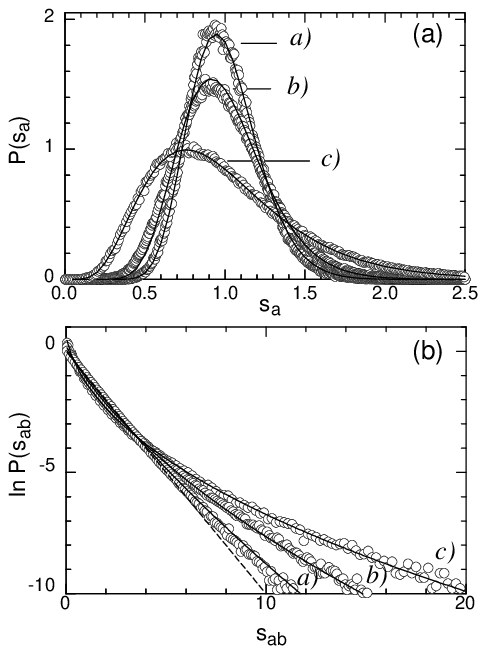}
\caption{Distribution functions of (a) the normalized total
transmission and (b) transmitted intensity, $P(s_a)$ and
$P(s_{ab})$, respectively, for three Polystyrene samples with
dimensions: $a)$ $d$=7.5 cm, $L=66.7$ cm; $b)$ $d$=5.0 cm,
$L=50.0$ cm; c) $d$=5.0 cm, $L=200.0$ cm. Solid lines represent
theoretical results obtained from Eqs.~(1) and (3), with measured values
of $2/3$var$(s_a)$ substituted for $g$. The dashed line in (b) is
a semi-logarithmic plot of the Rayleigh distribution,
$P(s_{ab})=\exp\left(-s_{ab}\right)$. From Refs. \cite{92,120}.}
\end{figure}

The theoretical expressions for the full distribution functions in
Eqs.~(1) and (3) are given as functions of $g$ for nonabsorbing samples.
We now consider the full transmission distributions in the
presence of absorption. Since the decreased value of $g$ due to
absorption does not lead to enhanced values of intensity
correlation but rather the reverse, it is not possible to utilize
the value of $g$ in Eqs.~(1) and (3). But since var$(s_a)=2/3g$,
$P(s_a)$ may be expressed as a function of var$(s_a)$. When
var$(s_a)$ is calculated for the three samples in Fig.~5, and
2/3var$(s_a)$ is substituted for $g$ in Eq.~(4), the resulting
curves plotted in the figure are in excellent agreement with the
measured distributions. The measured values of 2/3var$(s_a)$ may
be compared to the values of $g$ calculated without absorption
which were given above: 10.2 for sample $b$ instead of 9, and 3.06
for sample $c$ instead of 2.25. The measured values of
2/3var$(s_a)$ are larger when absorption is present, because
absorption reduces the weight of longer paths. The distribution of
Eq.~(4) gives the exponential tail
$P(s_a)\sim\exp[-2s_a/3$var$(s_a)]$ \cite{90,91}. These results
are in excellent agreement with measurements \cite{92}.

We find that var$(s_a)$ does not drop sharply with increasing
absorption as does the average transmission, but is only weakly
suppressed \cite{41}. This indicates that the broad distribution
of intensity and of total transmission, which is characteristic of
wave propagation in random systems, is preserved in absorbing
media. It is of interest therefore to examine the scaling of the
statistics of propagation. The scaling of var$(s_a)$ determined at
a number of frequencies is shown in Fig.~6 \cite{41}. Var$(s_a)$
increases exponentially once it becomes of the order of unity.
This is expected for a localization parameter, since, in the
absence of absorption, var$(s_a)\sim 1$, when $g\sim 1$. The
availability of a measurable localization parameter makes it
possible to definitively index the nature of propagation via a
parameter marking the closeness to the localization threshold. The
measurement of var$(s_a)$ makes it possible to probe the array of
factors that foster localization, including size, concentration,
and structural correlation.
\begin{figure}
\includegraphics [width=\columnwidth] {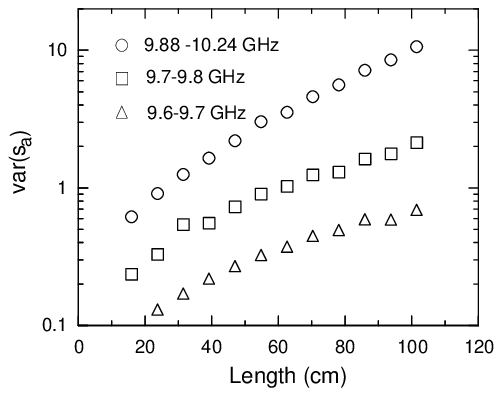}
\caption{Scaling of var$(s_a)$ in samples of alumina spheres.
Above a value of the order of unity, var$(s_a)$ increases
exponentially. From Ref. \cite{41}.}
\end{figure}

\section{Window of localization}

In the absence of absorption, localization can always be brought
about by increasing the length of a sample, while holdings its
cross-sectional area fixed \cite{2}. To induce localization in as
short a sample as possible, strongly scattering samples should be
used. Since the strength of scattering is peaked near Mie
resonances, it is useful to carry out measurements in samples
containing spherical scatterers at wavelengths comparable to the
diameter of the spheres. In order for the resonances not to be
washed out by correlated scattering among neighboring spheres,
samples with low density of spheres were studied. Microwave
measurements were made in quasi-1D samples of alumina spheres at a
volume fraction of $f=0.068$ \cite{55,121}. The alumina spheres
have a diameter of 0.95 cm and an index of refraction of $n=3.14$.
These are placed at the centers of 1.9-cm-diameter Styrofoam
spheres. The sample is contained in a 7.3-cm-diameter copper tube
with sample lengths up to $L=90$ cm. An ensemble of 5,000 random
sample configurations is created by rotating the tube between
subsequent field spectrum measurements. For each sample
configuration, the field spectrum yields the corresponding
frequency variation of the transmitted intensity $I_{ab}$. The
average intensity $\langle I_{ab}\rangle$ for $L=80$ cm samples,
shown in Fig.~7a, exhibits distinct drops near Mie resonances. The
resonant character of scattering is further indicated by the sharp
peaks in the average photon transit time, $\langle\tau\rangle$,
seen in Fig.~7b.
\begin{figure}
\includegraphics [width=\columnwidth] {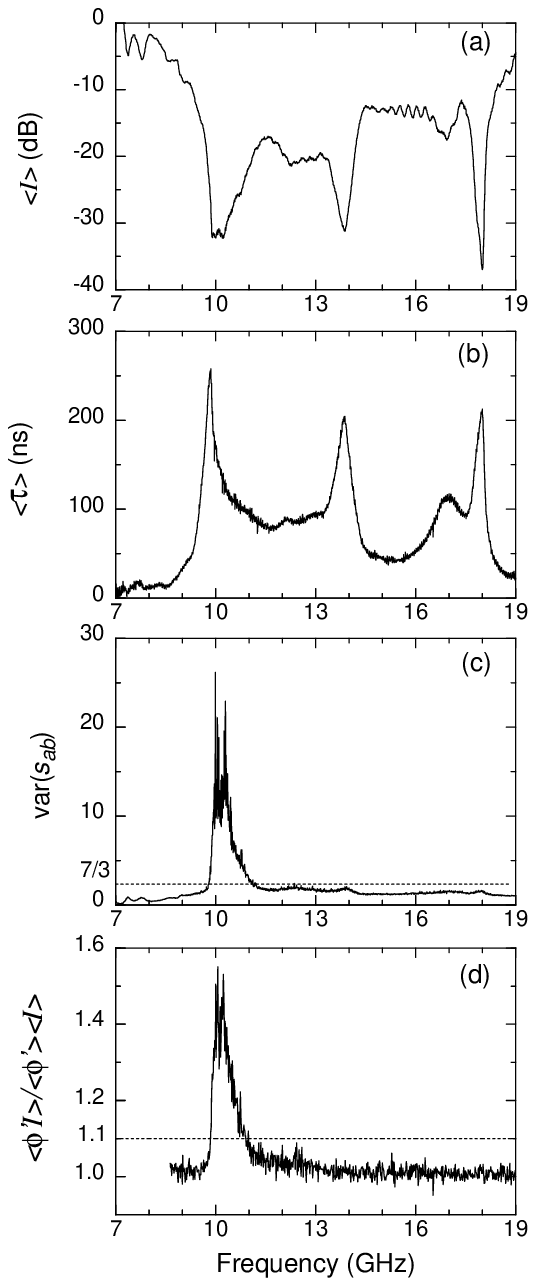}
\caption{(a) Average transmitted intensity, $\langle I\rangle$,
(b) average photon transit time, $\langle\tau\rangle$, (c)
variance of normalized transmitted intensity, var$(s_{ab})$, and
(d) dimensionless ratio,
$\langle\hat{\phi^{\prime}}\hat{I}\rangle=
\langle\phi^{\prime}I\rangle/\langle\phi^{\prime}\rangle\langle
I\rangle$, versus frequency in a quasi-1D alumina sample with
$L=80$ cm and alumina volume filling fraction $f=0.068$. The
dashed lines in (c) and (d) indicate the localization threshold.
From Refs.~\cite{55,133}.}
\end{figure}
The transit time or single-channel delay time is
given by $\tau_{ab}=s_{ab}d\phi_{ab}/d\omega$, where $\phi_{ab}$
is the phase accumulated by the field as it propagates through the
sample, from incident channel $a$ to outgoing channel $b$, and
$\omega$ is the angular frequency \cite{122,123}. Since low
transmission can be due to absorption and long dwell time can be
associated with microstructure resonances \cite{24,25,26}, the
average transmission and dwell time do not provide definitive
measures of the closeness to the localization threshold. This can
be obtained, however, from the measurement of var$(s_{ab})$, shown
in Fig.~7c. For diffusive waves obeying Rayleigh statistics,
var$(s_{ab})=1$. Lower values of var$(s_{ab})$, seen below 8.5
GHz, indicate a significant ballistic component in the transmitted
field, whereas higher values indicate the presence of substantial
long-range correlation. The horizontal dashed line in Fig.~7c
represents the localization threshold, var$(s_{ab})=7/3$, which
corresponds to var$(s_a)=2/3$. In alumina samples with $L=80$ cm,
the localization threshold is crossed in a narrow frequency range
above the first Mie resonance. Large fluctuations of intensity
reflected in measurements of var$(s_{ab})$ above the localization
threshold, seen in Fig.~7c, are a result of the greatly enhanced
tail of the intensity distribution for localized radiation
associated with the strong difference in transmission when a
sample is on or off resonance. Within the localization window,
var$(s_{ab})>7/3$, the probability distribution for intensity
becomes extraordinarily broad \cite{41}.

The strength of scattering cannot be substantially increased by
increasing the concentration of randomly positioned alumina
spheres. When the density of alumina spheres is increased
individual sphere resonances are washed out. This is shown in
Fig.~8a in measurements of $D$ as a function of volume filling
fraction, $f$, of nearly spherical alumina particles with
approximate diameter 0.95 cm and index of refraction $n=2.95$
\cite{27}. Measurements of the mean free path for $f=0.30$ are
shown in Fig.~8b \cite{124}. These results yield the transport
velocity shown in Fig.~8c, which is substantially reduced over a
broad spectral range in agreement with CPA calculations
\cite{125}.
\begin{figure}
\includegraphics [width=\columnwidth] {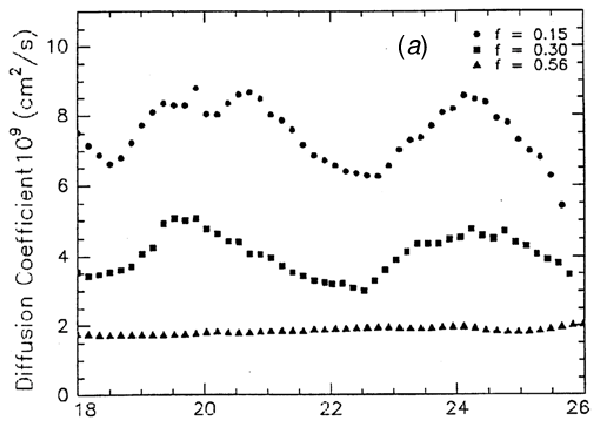}
\includegraphics [width=\columnwidth] {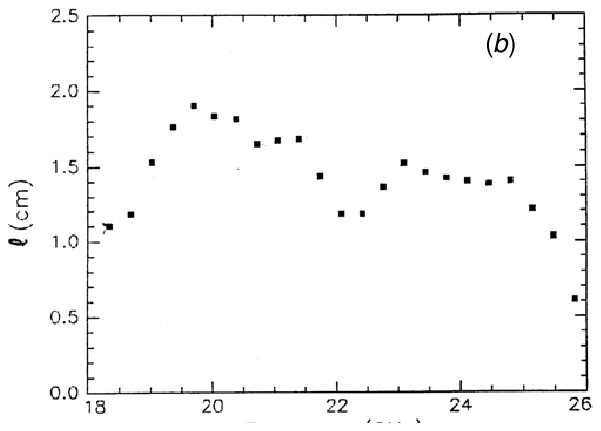}
\includegraphics [width=\columnwidth] {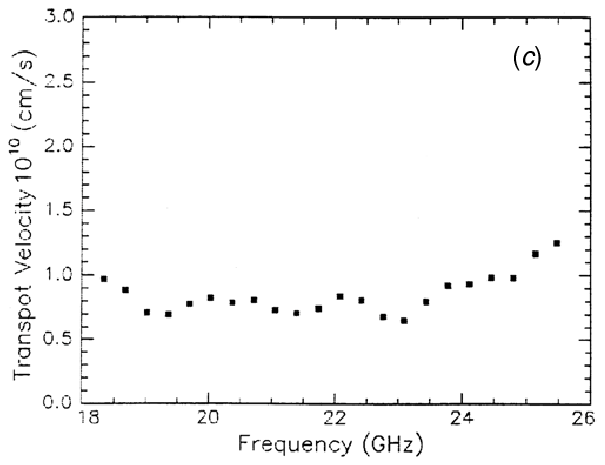}
\caption{(a) Frequency dependence of the diffusion coefficient for
mixtures of industrial-grade 0.95-cm-diameter alumina spheres
within hollow polypropylene spheres for various alumina 
filling fractions, $f$; (b) Transport mean free path for
sample with $f=0.30$; (c) Transport velocity for $f=0.30$ obtained
from (a) and (b), $v_E=3D/\ell$. From Refs.~\cite{27,124}.}
\end{figure}

\section{Delay time statistics}

In addition to the statistics of steady-state transmission, there
exists a rich statistics of dynamics, which is strongly correlated
to static aspects of propagation \cite{102,126,127}. For a
spectrally narrow pulse, the single-channel delay time is found to
be the derivative of the phase of the transmitted field with
frequency, $\tau_{ab}=s_{ab}d\phi_{ab}/d\omega\equiv\phi^{\prime}$
\cite{122,123}. The single-channel delay time can similarly be
defined for the reflected field. The interplay between the
single-channel delay time and transmission coefficient is
expressed in the joint probability distribution
$P(\phi^{\prime},I)$. The phase is discontinuous with a jump of
$\pi$ radians, as the null in intensity in the speckle pattern is
traversed \cite{128}, giving rise to strong fluctuations in
$\phi^{\prime}$ at low intensities. For a given value of intensity
$I$ the delay time $\phi^{\prime}$ is normally distributed with
normalized standard deviation,
$\Delta\phi^{\prime}/\langle\phi^{\prime}\rangle=\sqrt{Q/2I}$,
whereas the second-order cumulant correlator vanishes,
$\langle\phi^{\prime}I\rangle-\langle\phi^{\prime}\rangle\langle
I\rangle=0$. For, $\ell<L<\xi$, in the diffusive regime, in the
absence of absorption, the probability distribution of the delay
times is
$P(\hat{\phi^{\prime}}\equiv\phi^{\prime}/\langle\phi^{\prime}\rangle)=
(Q/2)[Q+(\hat{\phi^{\prime}}-1)^2]^{-3/2}$,
where $Q\approx 1$ in transmission \cite{129,130} and
$Q\approx\left(L/\ell\right)^2$ in reflection \cite{129,130,131}.
Absorption can be accounted for simply by a change in $Q$. This
agrees well with microwave \cite{129,130} and optical \cite{132}
measurements.

The statistics of dynamics of localized waves differ fundamentally
from those of diffusive waves since the transmission spectrum
includes narrow spikes, since $\delta<1$. Long delay times for
localized waves occur at peaks in transmission, associated with
resonant tunneling through localized states, so that
$\langle\phi^{\prime}I\rangle-\langle\phi^{\prime}\rangle\langle
I\rangle>0$. The correlator between $\phi^{\prime}$ and $I$ thus
provides a dynamical test of photon localization. The frequency
variation of the ratio $\langle\hat{\phi^{\prime}}\hat{I}\rangle=
\langle\phi^{\prime}I\rangle/\langle\phi^{\prime}\rangle\langle
I\rangle$ in a quasi-1D sample of alumina spheres with $L=80$ cm
is plotted in Fig.~7d \cite{133}. It is unity in the diffusive
limit and rises above unity as localization is approached.
$\langle\hat{\phi^{\prime}}\hat{I}\rangle$ and var$(s_a)$ both
rise appreciably only within the localization window in Fig.~7.
The localization threshold, at which var$(s_{ab})=7/3$,
corresponds to the condition
$\langle\hat{\phi^{\prime}}\hat{I}\rangle=1.1$, shown as the
dotted line. The conditional probability distribution
$P_I(\hat{\phi^{\prime}})$ given $I$ for localized waves in
transmission is found to exhibit an exponential fall-off, with an
asymmetry in the distribution, which increases with decreasing
$I$, and to have a normalized standard deviation of
$\Delta\phi^{\prime}/\langle\phi^{\prime}\rangle\propto I^{-0.25}$
\cite{133}. The distribution, $P(\phi^{\prime})$, of the
single-channel delay time is given by random-matrix theory
\cite{131}. In the localized regime, it is found to have a
universal quadratic tail,
$P(\phi^{\prime})\propto\left|\phi^{\prime}\right|^{-2}$
\cite{134,135,136}. In samples of the finite length $L$, this
algebraic tail eventually crosses over to a log-normal tail, at
exponentially long delay times,
$|\phi^{\prime}|>\tau_s\exp(L/\xi)$, where $\tau_s$ is the
scattering time \cite{103,104,137,138}. There is an interesting
dynamic signature of localization in reflection: the distribution
$P(\phi^{\prime})$ differs in the two cases, $a=a^{\prime}$ and
$a\neq a^{\prime}$. The maximal value of $P(\phi^{\prime})$ in the
case $a=a^{\prime}$  is greater than that for $a\neq a^{\prime}$
by a factor close to $\sqrt{2}$ \cite{131}. The effect appears
only in the localized regime, as shown in Fig.~9.
\begin{figure}
\includegraphics [width=\columnwidth] {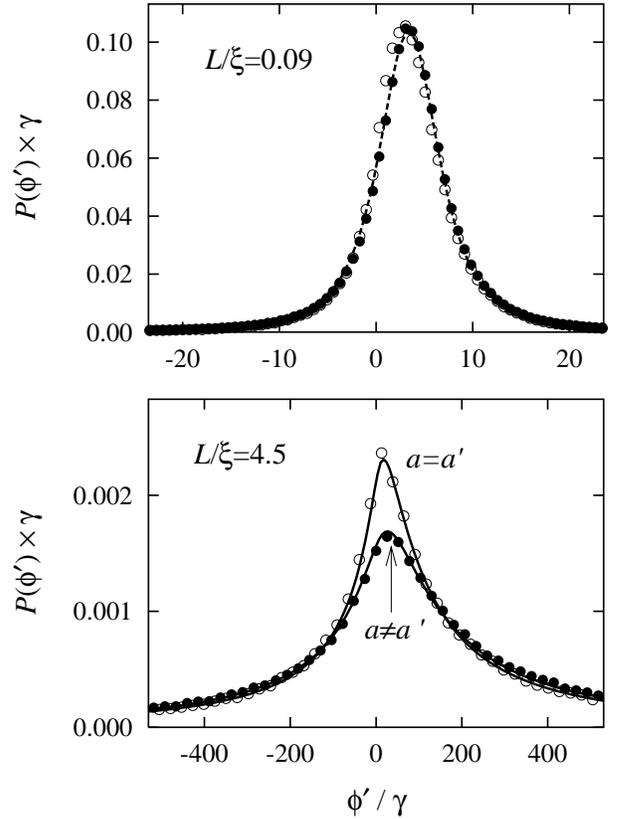}
\caption{Calculations of distribution of the single-channel
reflection delay time in the diffusion (top panel) and localized
(bottom panel) regimes \cite{131}. The maximal value of
$P\left(\phi^{\prime}\right)$ in the case $a=a^{\prime}$ of equal
excitation and reflection modes (open circles) is greater than
that for $a\neq a^{\prime}$  (solid circles) by a factor close to
$\sqrt{2}$. The effect appears only in the localized regime. (with
kind permission of the authors)}
\end{figure}
While there is a complete theory of the statistics of dynamics in
reflection, in transmission only the 1D case, $N=1$, has been
solved exactly \cite{139}. In this case, the transmission delay
time is found to be the mean of reflection delay times of the both
ends of the waveguide. For $N>1$, an approximate solution was
found \cite{139}, which is shown in Fig.~10a, and compared to
results of numerical simulations. Microwave measurements of
$P(\phi^{\prime})$ within the localization window were carried out
in ensembles of alumina samples of increasing length \cite{133}.
The distributions of the normalized single-channel delay time in
samples with $L=25$ and 90 cm are shown in Fig.~10b. The values of
var$(s_a)$ in these samples are 1.0 and 7.1, and the
ensemble-averaged values of $\phi^{\prime}$ are
$\langle\phi^{\prime}\rangle=21$ and 122 ns, respectively. As
Fig.~10b suggests, $P(\phi^{\prime})$ becomes more asymmetric as
$L$ increases, and reaches its peak at values of
$\hat{\phi^{\prime}}$ further below its average value of unity.
The measured distributions of Fig.~10b appear to be similar to the
calculated distributions in Fig.~10a, though a closer comparison
reveals a discrepancy in the tail that may be due to the presence
of absorption in alumina samples.
\begin{figure}
\includegraphics [width=\columnwidth] {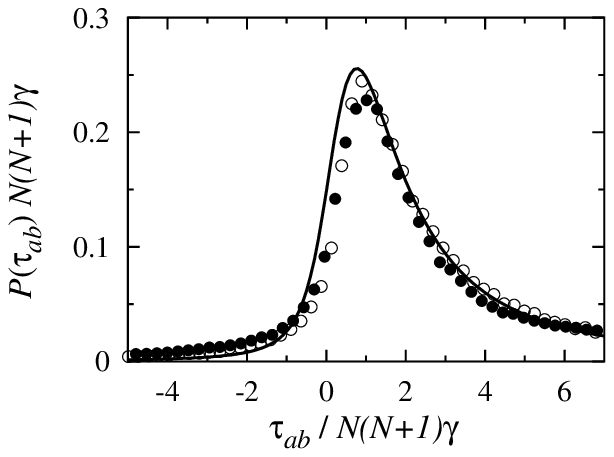}
\includegraphics [width=\columnwidth] {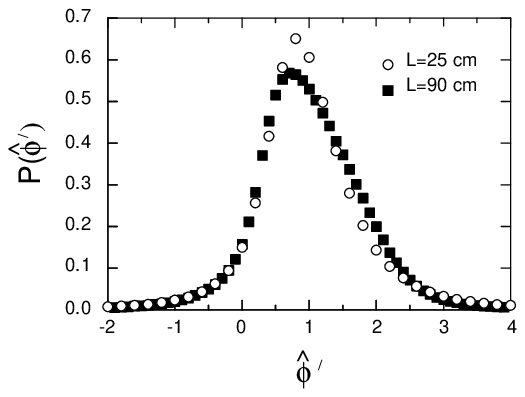}
\caption{(a) Distribution of the single-channel transmission delay
time obtained from numerical simulations of random scattering in
planar waveguides with $N=2$ (open circles) and $N=30$ (filled
circles) channels. The solid line is the analytical prediction,
independent of $N$ \cite{139}. (with kind permission of the author);
(b) Probability distribution of the normalized delay time,
$P(\hat{\phi^{\prime}})$, in alumina samples with $L=25$ (circles)
and 90 cm (squares) \cite{133}.}
\end{figure}

The averaged delay time is found by summing the single-channel
delay time weighted by the intensity,
$W_{ab}=I_{ab}\phi^{\prime}_{ab}$. When summed over all input and
output modes, $W_{ab}$ relates directly to a fundamental dynamic
quantity in condense matter, namely the number of microstates per
unit frequency interval $d\omega$ inside the scattering medium,
$N(\omega)$, \cite{140}
\begin{equation}
{1\over\pi}\Sigma_{ab}^{2N} I_{ab}\phi^{\prime}_{ab}=N(\omega).
\label{}
\end{equation}
The summation runs over $N$ channels in both reflection and
transmission. Though the density of states in an open (scattering)
system is ill-defined, the l.h.s. of the Eq.~(5) is well defined
and proportional to $\int dr\left(|\psi_{\omega}(r)|^2-1\right)$,
\cite{140} which equals the stored electromagnetic energy
\cite{141}. In the diffusive limit of Gaussian field statistics,
the distribution of $W=W_{ab}$ is a double-sided exponential
\cite{129,130}. In a 1D sample with $L>\xi$, an algebraic decay
with exponent -4/3 was calculated for $P(W)$ using a model of
resonant transport via localized states \cite{135}. The
probability distribution of the normalized weighted delay time,
$P\left(\hat{W}\equiv W/\langle W\rangle\right)$, for localized
waves in a quasi-1D alumina samples with $L=90$ cm is shown in
Fig.~11 \cite{133}. The distribution $P(\hat{W})$ is extraordinary
broad, reflecting the enhanced probability of long dwell times and
the increased variance of dwell times at larger values of the
intensity for localized waves. It can be well approximated by a
double-sided stretched exponential to the power of 1/3 (Fig.~11).
The affect of absorption on the form of the measured distributions
was not studied.
\begin{figure}
\includegraphics [width=\columnwidth] {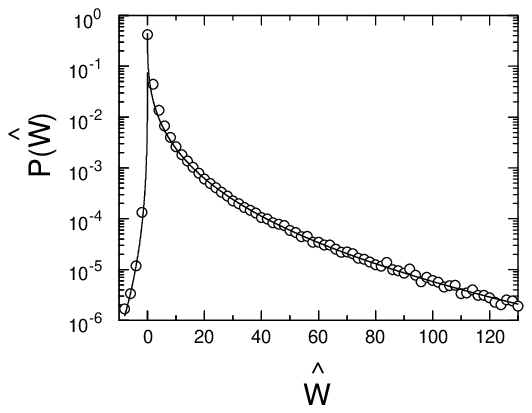}
\caption{Probability distribution of the normalized weighted delay
time, $P(\hat{W})$, for localized waves (circles) in alumina
sample with $L=90$ cm. The solid line is the model distribution,
$P(\hat{W})=a\exp\left(-b|\hat{W}|^{1/3}\right)$, with $a=0.44$
and $b=2.42$ for $\hat{W}>0$, and $a=0.07$ and $b=5.50$ for
$\hat{W}<0$. From Ref.~\cite{133}.}
\end{figure}

Notice that, though less probable, negative ``delay" times for the
multimode case are observed and are also allowed by scattering
theory \cite{142,143}. Negative delay times arise as a result of
pulse reshaping \cite{123}. This is in contrast to the ``proper"
delay times defined as the eigenvalues of the Wigner-Smith matrix
$Q=-iS^*\cdot\partial S/\partial\omega$ and its trace
Tr$\,Q=\sum_{ab}W_{ab}=\pi N(\omega)=\tau_H$, called the
Heisenberg time $\tau_{H}$; the channel-average
$\tau_{H}/2N=\tau_{W}$ is associated with the Wigner-Smith delay
time $\tau_{W}$. Substantial theoretical effort in mesoscopic
physics has focused on the probability distributions of the
``proper" delay times and the Wigner-Smith delay time
\cite{144,145,146,147}, whose measurements require knowledge of
the entire Wigner-Smith matrix. The joint distribution of the
inverse ``proper" delay times, $\gamma_n=1/\tau_n$, was found to
be the Laguerre ensemble of random-matrix theory \cite{145}, in
which the correlation functions of $\tau_n$'s are given in terms
of Laguerre polynomials, whereas $P(\tau_W)$ was found to follow
universal distributions with algebraic decays in the localized
\cite{134,135,136} and critical \cite{148,149} regimes,
respectively. It has also been shown that the statistics of delay
times relates to the statistics of the eigenfunction intensities
and may be used to probe eigenfunction fluctuations in closed
mesoscopic samples of any spatial dimension \cite{150,151}.

\section{Statistics of pulsed transmission}

In addition to probing the time evolution of average transmission
and its increasing suppression, and the statistics of delay time
for incident pulses, it is possible to measure the spatial and
polarization correlation functions and the full distribution of
the field and intensity at any value of $t$ \cite{152}. The
cumulant correlation function with displacement and polarization
rotation of intensity of the transmitted wave normalized to its
ensemble average value for an incident pulse of linewidth $\sigma$
at time $t$ has the form,
\begin{equation}
C=F+\kappa_{\sigma}(t)\left(1+F\right), \label{}
\end{equation}
where $F$ is the complex square of the field correlation function.
The parameter $\kappa_{\sigma}(t)$ establishes the degree of
correlation at a delay time $t$ from the center of an exciting pulse
of bandwidth $\sigma$, and is evidently the fractional intensity
correlation at a point at which $F$ vanishes. This is the same
form as the corresponding steady-state correlation functions with
the degree of steady state correlation $\kappa_0$ instead of
$\kappa_{\sigma}(t)$ \cite{153,154}. The field correlation
function is found not to be a function of time, while the degree
of intensity correlation $\kappa_{\sigma}(t)$ varies with time and
depends upon the pulse bandwidth \cite{152}. This is seen in the
polarization measurements shown in Fig.~12 carried out in an
ensemble of 12,000 configurations of low density alumina spheres
at $L=90$ cm over the frequency range 16.95-17.05 GHz near the
peak of the fourth Mie resonance with spectral width $\sigma=5$
MHz \cite{152}. For the conditions given above, $\kappa_0=0.29$.
The dynamic probability distribution of normalized intensity is
found to depend only upon $\kappa_{\sigma}(t)$, which is equal to
the variance of the normalized total transmission at a fixed delay
time, var$(s_a(t))$. The intensity distribution at a fixed delay
time is given by Eq.~(1) with $g=2/3\kappa$ and closely
corresponds to the measured intensity distribution, even for
$\kappa\gg 1$. Values of $\kappa_{\sigma}(t)$ much greater than
$\kappa_{0}$  are found at long delay times. For an incident pulse
width $\sigma=1$ MHz in a sample with $L=61$ cm, in the frequency
range 9.95-10.05 GHz, just above the first Mie resonance, at
$t=740$ ns, we find $\kappa=9.55$.
\begin{figure}
\includegraphics [width=\columnwidth] {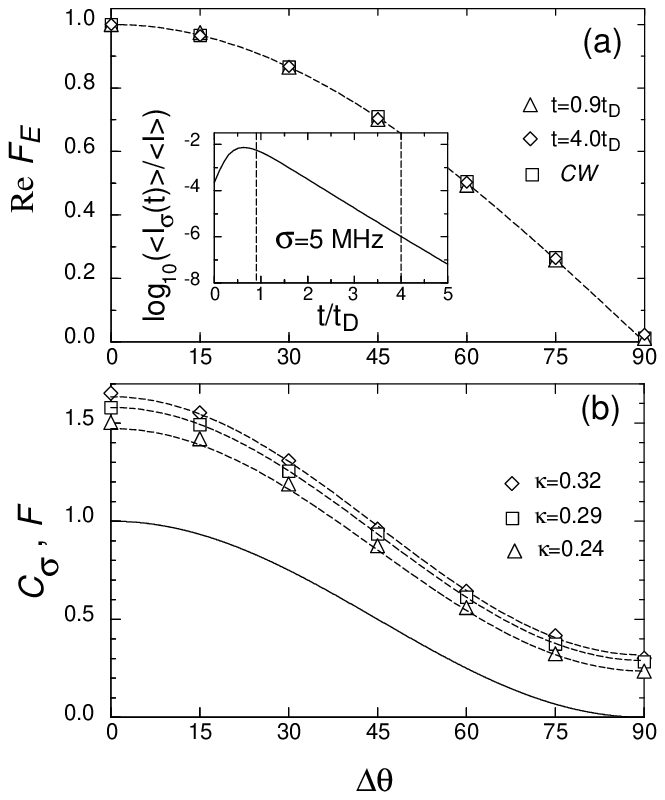}
\caption{(a) Real part of the field correlation function,
$F_E(\Delta\theta)$, and (b) intensity correlation function,
$C_{\sigma}(\Delta\theta)$, with polarization rotation of the
transmitted wave at the two delay times following pulsed
excitation by a Gaussian pulse with spectral width $\sigma=5$ MHz
and for monochromatic excitation (CW). The dashed curves are (a)
$F_E(\Delta\theta)=\cos(\Delta\theta)$ and (b)
$C_{\sigma}=F(\Delta\theta)+\kappa_{\sigma}(t)\left[F(\Delta\theta)+1\right]$,
where $F(\Delta\theta)=|F_E(\Delta\theta)|^2$ with the values of
$\kappa_{\sigma}(t)$ indicated. The solid curve is
$F(\Delta\theta)$. The logarithm of the average pulsed
transmission through the alumina sample for $\sigma=5$ MHz,
normalized by the average steady-state transmitted intensity, is
shown in the inset. The delay times at which correlation is
presented in the figure are indicated by vertical dashed lines.
From Ref.~\cite{152}.}
\end{figure}

\section{Field statistics}

The isotropic approximation of random matrix theory is equivalent
to the assumption that the field distribution in a given
configuration is a circular Gaussian \cite{155}. This is
demonstrated by measurements of the microwave field in random
media, which show that the probability distribution, $P(E)$, of
the field normalized to the ensemble average of the field
amplitude,
$E=t_{ab}/\sqrt{\langle|t_{ab}|^2\rangle}=r+\sqrt{-1}i$, is given
by substituting $r^2+i^2$ for $s_{ab}$ in Eq.~(1) \cite{156}. The
probability distribution for the transmitted field for a
sub-ensemble with a fixed value of $s_a$ is thus Gaussian. Because
the field in samples with a fixed value of $s_a$ is a Gaussian
random process, the field normalized to its average value in a
given configuration and the total transmission are statistically
independent. The field correlation function may therefore be
written as a product of correlators of the normalized Gaussian
field and the square root of the total transmission. Since the
field correlation function with frequency shift is the Fourier
transform of the time-of-flight distribution \cite{29,30}, pulse
dynamics, which shows progressive suppression of transport, may be
expressed in terms of the spectral correlation function of the
normalized Gaussian field and the square root of total
transmission. We expect that the correlation function of the
Gaussian field will not be affected by mesoscopic fluctuations. On
the other hand, the correlation function of the square root of the
total transmission reflects mesoscopic fluctuations. In this way
the dynamics of the field with the observed suppression of
transmission in the time domain due to the growing impact of
localization is directly tied to mesoscopic fluctuations.

\section{Conclusion}

Because the character of wave transport depends upon the proximity
to the localization threshold, signatures of localization, and of
the approach to localization, can be found whenever statistical
measurements can be made in ensembles of random sample
realizations. A large number of statistically different sample
configurations may arise as a result of constant motion of
colloidal systems, or in powdered samples, or in collections of
scattering particles that can be tumbled. They may also be created
by illuminating different parts of a large slab or by tuning the
frequency of a monochromatic source through a spectrum much wider
than the correlation frequency. The probability distributions of
intensity and total transmission in steady-state and pulsed
measurements can be described in terms of the variance of
normalized total transmission, which is equal to the degree of
intensity correlation, var$(s_a)=\kappa$. The variation of
$\kappa$ with delay time directly charts the increasing impact of
localization and is independent of absorption. The decrease of
fluctuations and correlation with absorption in steady-state
measurements reflects the reduced weight with increasing
absorption of long trajectories associated with strong
correlation.

Absorption reduces both steady-state and pulsed transmission. The
influences of absorption and weak localization can be isolated in
the time domain. Absorption reduces the intensity of pulsed
transmission exponentially. This reduction can be cancelled in the
analysis of data by multiplying by the corresponding growing
exponential, to yield the transmitted pulse that would be observed
in the absence of absorption. The role of localization in averaged
transmission at a given time delay is thus displayed directly. In
contrast, absorption and localization affects are mixed in
steady-state transmission, and either can lead to an exponential
decrease of transmission with sample thickness.

The suppression of the spread of a beam within a sample also gives
an indication of localization. This can be observed in the
broadening of the peak and in the reduction of the coherent
backscattering enhancement factor below 2. Other signatures of
localization are observed in measurements of the statistics of the
single channel delay time and its correlation with the intensity.
Given the many signatures of localization, the state of
propagation can be unambiguously determined even in the presence
of absorption.

\section{Acknowledgements}

We gratefully acknowledge stimulating discussion with Z.Q. Zhang,
P. Sebbah, A.A. Lisyansky, B. Hu, B.A. van Tiggelen, and J.H. Li.
This work was supported by the National Science Foundation under
grants DMR0205186 and DMR0538350.


\begin{thebibliography}{999}

\bibitem{1} P.W. Anderson, Phys. Rev. {\bf 109}, 1492 (1958).
\bibitem{2} D.J. Thouless, Phys. Rev. Lett. {\bf 39}, 1167 (1977).
\bibitem{3} E. Abrahams, P.W. Anderson, D.C. Licciardello, and T.V. Ramakrishnan, Phys. Rev. Lett. {\bf 42}, 673 (1979).
\bibitem{4} P.W. Anderson, D.J. Thouless, E. Abrahams, and D.S. Fisher, Phys. Rev. B {\bf 22}, 3519 (1980).
\bibitem{5} {\it Mesoscopic Phenomena in Solids}, eds. B.L. Altshuler, P.A. Lee, and R.A. Webb, (Elsevier, Amsterdam, 1991).
\bibitem{6} A. Ioffe and A.R. Regel, Prog. Semicond. {\bf 4}, 237 (1960).
\bibitem{7} M.E. Gertsenshtein and V.B. Vasil'ev, Theor. Probab. Appl. {\bf 4}, 391 (1959).
\bibitem{8} G. Papanicolaou, J. Appl. Math. {\bf 21}, 13 (1971).
\bibitem{9} P.L. Sulem and U. Frisch, J. Plasma Phys. {\bf 8}, 217 (1972).
\bibitem{10} S. John, Phys. Rev. Lett. {\bf 53}, 2169 (1984).
\bibitem{11} P.W. Anderson, Philos. Mag. {\bf 52}, 505 (l985).
\bibitem{12} {\it Scattering and Localization of Classical Waves in Random Media},
ed. P. Sheng, (World Scientific Press, Singapore, l990).
\bibitem{13} S. John, Physics Today {\bf 44}, (May, 1991).
\bibitem{14} A. Lagendijk, in {\it Current Trends in Optics}, ed. J.C. Dainty (Academic, London, 1994).
\bibitem{15} R.A. Webb, S. Washburn, C.P. Umbach, and R.B. Laibowitz, Phys. Rev. Lett. 54, 2696 (1985).
\bibitem{16} B.L. Altshuler and D.E. Khmelnitskii, JETP Lett. {\bf 42}, 359 (1985).
\bibitem{17} P.A. Lee and A.D. Stone, Phys. Rev. Lett. {\bf 55}, 1622 (1985).
\bibitem{18} M.P. Van Albada and A. Lagendijk, Phys. Rev. Lett. {\bf 55}, 2692 (1985).
\bibitem{19} P.E. Wolf and G. Maret, Phys. Rev. Lett. {\bf 55}, 2696 (1985).
\bibitem{20} E. Akkermans, P.E. Wolf, and R. Maynard, Phys. Rev. Lett. {\bf 56}, 1471 (1986).
\bibitem{21} D.S. Wiersma, M.P. van Albada, B.A. van Tiggelen, and A. Lagendijk, Phys. Rev. Lett. {\bf 74}, 4193 (1995).
\bibitem{22} J.H. Li and A.Z. Genack, Phys. Rev. E {\bf 49}, 4530 (1994).
\bibitem{23} A.Z. Genack, Phys. Rev. Lett. {\bf 58}, 2043 (1987).
\bibitem{24} M.P. van Albada, B.A. van Tiggelen, A. Lagendijk, and A. Tip, Phys. Rev. Lett. {\bf 66}, 3132 (1991).
\bibitem{25} A. Lagendijk and B.A. van Tiggelen, Phys. Rep. {\bf 270}, 143 (1996).
\bibitem{26} D. Livdan and A.A. Lisyansky, Phys. Rev. B {\bf 53}, 14843 (1996).
\bibitem{27} A.Z. Genack, J.H. Li, N. Garcia, and A.A. Lisyansky, in {\it Photonic Band Gaps and Localization},
ed. C.M. Soukoulis, (Plenum, New York, 1993).
\bibitem{28} G.H. Watson, Jr., P.A. Fleury, and S.L. McCall, Phys. Rev. Lett. {\bf 58}, 945 (1987).
\bibitem{29} J.M. Drake and A.Z. Genack, Phys. Rev. Lett. {\bf 63}, 259 (1989).
\bibitem{30} A.Z. Genack and J.M. Drake, Europhys. Lett. {\bf 11}, 331 (1990).
\bibitem{31} K.M. Yoo, F. Liu, and R.R. Alfano, Phys. Rev. Lett. {\bf 64}, 2647 (1990); {\it ibid} {\bf 65}, 2210 (1990).
\bibitem{32} R.H.J. Kop, P. deVries, R. Sprik, and A. Lagendijk, Phys. Rev. Lett. {\bf 79}, 4369 (1997).
\bibitem{33} A.A. Chabanov, Z.Q. Zhang, and A.Z. Genack, Phys. Rev. Lett. {\bf 90}, 203903 (2003).
\bibitem{34} B. Shapiro, Phys. Rev. Lett. {\bf 57}, 2168 (1986).
\bibitem{35} M.P. Van Albada, J.F. de Boer, and A. Lagendijk, Phys. Rev. Lett. {\bf 64}, 2787 (1990).
\bibitem{36} P. Sebbah, R. Pnini, and A.Z. Genack, Phys. Rev. E {\bf 62}, 7348 (2000).
\bibitem{37} G. Maret and P.E. Wolf, Zeitschrift fur Physik B {\bf 65}, 409 (1987).
\bibitem{38} D.J. Pine, D.A. Weitz, P.M. Chaikin, and E. Herbolzheimer, Phys. Rev. Lett. {\bf 60}, 1134 (1988).
\bibitem{39} N. Garcia and A.Z. Genack, Phys. Rev. Lett. {\bf 63}, 1678 (1989).
\bibitem{40} J.B. Pendry, Nature {\bf 342}, 223 (1989).
\bibitem{41} A.A. Chabanov, M. Stoytchev, and A.Z. Genack, Nature {\bf 404}, 850 (2000).
\bibitem{42} M.Ya. Azbel, Solid State Commun. {\bf 45} 527 (1983).
\bibitem{43} M.Ya. Azbel and P. Soven, Phys. Rev. B {\bf 27}, 831 (1983).
\bibitem{44} V. D. Freilikher and S. A. Gredeskul, in {\it Progress in
Optics},Volume 30, ed. by E. Wolf, (Elsevier, Amsterdam, 1992), p.
137.
\bibitem{45} I.M. Lifshits and V.Ya. Kirpichenkov, Sov. Phys. JETP {\bf 50}, 499 (1979).
\bibitem{46} J.B. Pendry, J. Phys. C {\bf 20}, 733 (1987).
\bibitem{47} J.B. Pendry, Adv. Phys. {\bf 43}, 461 (1994).
\bibitem{48} A. MacKinnon and B. Kramer, Phys. Rev. Lett. {\bf 47}, 1546 (1981).
\bibitem{49} J.B. Pendry, Nature {\bf 351}, 438 (1991).
\bibitem{50} M. Stoytchev and A.Z. Genack, Phys. Rev. B {\bf 55}, R8617 (1997).
\bibitem{51} U. Kuhl and H.-J. Stockmann, Physica E {\bf 9}, 384 (2001).
\bibitem{52} P. Sebbah, B. Hu, J. Klosner, and A.Z. Genack, submitted for publication.
\bibitem{53} R. Dalichaouch, J.P. Armstrong, S. Schultz, P.M. Platzman, and S.L. McCall, Nature {\bf 354}, 53 (1991).
\bibitem{54} C. Vanneste and P. Sebbah, Phys. Rev. Lett. {\bf 87}, 183903 (2001).
\bibitem{55} A.A. Chabanov and A.Z. Genack, Phys. Rev. Lett. {\bf 87}, 153901 (2001).
\bibitem{56} M.V. Berry and S. Klein, Eur. J. Phys. {\bf 18}, 222 (1997).
\bibitem{57} V. Milner and A.Z. Genack, Phys. Rev. Lett. {\bf 94}, 073901 (2005).
\bibitem{58} J. Bertolotti, S. Gottardo, D.S. Wiersma, M. Ghulinyan, and L. Pavesi, Phys. Rev. Lett. {\bf 94}, 113903 (2005).
\bibitem{59} S. John, Phys. Rev. Lett. {\bf 58}, 2486 (1987).
\bibitem{60} V.P. Bykov, Sov. Phys. JETP {\bf 35}, 269 (1972).
\bibitem{61} E. Yablonovitch, Phys. Rev. Lett. {\bf 58}, 2059 (1987).
\bibitem{62} A. Cohen, Y. Roth, and B. Shapiro, Phys. Rev. B {\bf 38}, 12125 (1988).
\bibitem{63} L.I. Deych, D. Zaslavsky, and A.A. Lisyansky, Phys. Rev. Lett. {\bf 81}, 5390 (1998).
\bibitem{64} L.I. Deych, A.A. Lisyansky, and B.L. Altshuler, Phys. Rev. Lett. {\bf 84}, 2678 (2000).
\bibitem{65} L.I. Deych, M.V. Erementchouk, and A.A. Lisyansky, Phys. Rev. Lett. {\bf 90}, 126601 (2003).
\bibitem{66} L.I. Deych, M.V. Erementchouk, A.A. Lisyansky, and B.L. Altshuler, Phys. Rev. Lett. {\bf 91}, 096601 (2003).
\bibitem{67} L.I. Deych, A.A. Lisyansky, and B.L. Altshuler, Phys. Rev. B {\bf 64}, 224202 (2001).
\bibitem{68} Y. Asada, K. Slevin, T. Ohtsuki, L.I. Deych, A.A. Lisyansky, and B.L. Altshuler, J. Phys. Soc. Jap. {\bf 72A}, 173 (2003).
\bibitem{69} K. Slevin, Y. Asada, and L.I. Deych, Phys. Rev. B {\bf 70}, 054201 (2004)
\bibitem{70} J. Prior, A. M. Somoza, and M. Ortuno, Phys. Rev. B {\bf 72}, 024206 (2005)
\bibitem{71} A.D. Stone, P.A. Mello, K.A. Muttalib, and J.L. Pichard, in {\it Mesoscopic Phenomena in Solids},
eds. B.L. Altshuler, P.A. Lee, and R.A. Webb, (Elsevier,
Amsterdam, 1991).
\bibitem{72} C.W.J. Beenakker, Rev. Mod. Phys. {\bf 69}, 731 (1997).
\bibitem{73} M.C.W. van Rossum and Th.M. Nieuwenhuizen, Rev. Mod. Phys. {\bf 71}, 313 (1999).
\bibitem{74} A.Z. Genack and A.A. Chabanov, in {\it Wave and Imaging through Complex Media},
ed. by P. Sebbah, (Kluwer, Dordrecht, 2001).
\bibitem{75} R. Landauer, Philos. Mag. {\bf 21}, 863 (1970).
\bibitem{76} E.N. Economou and C.M. Soukoulis, Phys. Rev. Lett. {\bf 46}, 618 (1981).
\bibitem{77} D.S. Fisher and P.A. Lee, Phys. Rev. B {\bf 23}, R6851 (1981).
\bibitem{78} S.K. Cheung, X. Zhang, Z.Q. Zhang, A.A. Chabanov, and A.Z. Genack, Phys. Rev. Lett. {\bf 92}, 173902 (2004).
\bibitem{79} A.Z. Genack, Europhys. Lett. {\bf 11}, 733 (1990).
\bibitem{80} M.J. Stephen and G. Cwilich, Phys. Rev. Lett. {\bf 59}, 285 (1987).
\bibitem{81} P.A. Mello, Phys. Rev. Lett. {\bf 60}, 1089 (1988).
\bibitem{82} P.A. Mello, E. Akkermans, and B. Shapiro, Phys. Rev. Lett. {\bf 61}, 459-462 (1988).
\bibitem{83} S. Feng, C. Kane, P.A. Lee, and A.D. Stone, Phys. Rev. Lett. {\bf 61}, 834 (1988).
\bibitem{84} A.Z. Genack, N. Garcia, and W. Polkosnik, Phys. Rev. Lett. {\bf 65}, 2129 (1990).
\bibitem{Wolfle} D. Vollhardt and P. Wolfle, Phys. Rev. B {\bf 22}, 4666 (1980).
\bibitem{85} N. Garcia and A.Z. Genack, Phys. Rev. Lett. {\bf 66}, 1850 (1991).
\bibitem{86} A.Z. Genack and N. Garcia, Phys. Rev. Lett. {\bf 66}, 2064 (1991).
\bibitem{87} D.S. Wiersma, P. Bartolini, A. Lagendijk, and R. Righini, Nature {\bf 390}, 671 (1997).
\bibitem{88} F. Scheffold, R. Lenke, R. Tweer, and G. Maret, Nature {\bf 389}, 206 (1999).
\bibitem{89} R.L. Weaver, Phys. Rev. B {\bf 47}, 1077 (1993).
\bibitem{90} E. Kogan and M. Kaveh, Phys. Rev. B {\bf 52}, R3813 (1995).
\bibitem{91} Th.M. Nieuwenhuizen and M.C.W. van Rossum, Phys. Rev. Lett. {\bf 74}, 2674 (1995).
\bibitem{92} M. Stoytchev and A.Z. Genack, Phys. Rev. Lett. {\bf 79}, 309 (1997).
\bibitem{93} P.M. Morse and H. Feshbach, {\it Methods of Theoretical Physics}, (McGraw-Hill, New York, 1953).
\bibitem{94} A. Lagendijk, R. Vreeker, and P. DeVries, Phys. Lett. A {\bf 136}, 81 (1989).
\bibitem{95} J. X. Zhu, D. J. Pine, and D. A. Weitz, Phys. Rev. A {\bf 44}, 3948 (1991).
\bibitem{96} K. Bodland, J. Sound Vib. {\bf 73}, 19 (1980).
\bibitem{97} F. Kawakami and K. Yamiguchi, J. Acoust. Soc. Am. {\bf 80}, 543 (1986).
\bibitem{98} J. Burkhardt and R.L. Weaver, J. Acoust. Soc. Am. {\bf 100}, 320 (1996).
\bibitem{99} O.L. Lobkis, R.L. Weaver, and I. Rozhkov, J. Sound Vib. {\bf 237}, 281 (2000).
\bibitem{100} E. Doron, U. Smilansky, and A. Frenkel, Phys. Rev. Lett. {\bf 65}, 3072 (1990).
\bibitem{101} H. Alt, H.-D. Graf, H.L. Harney, R. Hofferbert, H. Lengeler, A. Richter, P. Schardt, and
H.A.Weidenmuller, Phys. Rev. Lett. {\bf 74}, 62 (1995).
\bibitem{102} B.L. Altshuler, V.E. Kravtsov, and I.L. Lerner, in {\it Mesoscopic Phenomena in Solids},
eds. B.L. Altshuler, P.A. Lee, and R.A. Webb, (Elsevier,
Amsterdam, 1991).
\bibitem{103} B.A. Muzykantskii and D.E. Khmelnitskii, Phys. Rev. B {\bf 51}, 5480 (1995).
\bibitem{104} A.D. Mirlin, Phys. Rep. {\bf 326}, 259 (2000).
\bibitem{105} K.B. Efetov, Adv. Phys. {\bf 32}, 53 (1983).
\bibitem{106} S.E. Skipetrov and B.A. van Tiggelen, Phys. Rev. Lett. {\bf 92}, 113901 (2004).
\bibitem{107} S.K. Cheung and Z.Q. Zhang, cond-mat/0509381 (2005).
\bibitem{108} S.E. Skipetrov and B.A. van Tiggelen, cond-mat/05087269 (2005).
\bibitem{110} N. Lawandy, R. Balachandran, A. Gomes, and E. Sauvain, Nature {\bf 368}, 436 (1994).
\bibitem{109} A.Z. Genack and J.M. Drake, Nature {\bf 368}, 400 (1994).
\bibitem{111} H. Cao, Y.G. Zhao, S.T. Ho, E.W. Seelig, Q.H. Wang, and R.P.H. Chang, Phys. Rev. Lett. {\bf 82}, 2278 (1999).
\bibitem{112} S.V. Frolov, Z.V. Vardeny, and K. Yoshino, Phys. Rev. B {\bf 57}, 9141 (1999).
\bibitem{113} X.Jiang and C.M. Soukoulis, Phys. Rev. Lett. {\bf 85}, 70 (2000).
\bibitem{114} V.M. Apalkov, M.E. Raikh, and B. Shapiro, Phys. Rev. Lett. {\bf 89}, 016802 (2002).
\bibitem{115} A. Yamilov and H. Cao, Phys. Rev. A {\bf 69}, 031803(R) (2004).
\bibitem{116} V.I. Kopp, B. Fan, H.K.M. Vithana, and A.Z. Genack, Opt. Lett. {\bf 23}, 1707 (1998).
\bibitem{117} P.V. Shibaev, V.I. Kopp, and A.Z. Genack, Jour. Phys. Chem. B {\bf 107}, 6961 (2003).
\bibitem{118} J. Schmidtke, W. Stille, and H. Finkelmann, Phys. Rev. Lett. {\bf 90}, 083902 (2003).
\bibitem{119} V.I. Kopp and A.Z. Genack, Phys. Rev. Lett. {\bf 89}, 033901 (2002).
\bibitem{120} M. Stoytchev and A.Z. Genack, Opt. Lett. {\bf 24}, 262 (1999).
\bibitem{121} A.A. Chabanov and A.Z. Genack, in {\it Wave Scattering in Complex Media: from Theory to Applications},
eds. B.A. van Tiggelen and S. Skipetrov, (Kluwer Academic,
Dordrecht, 2003).
\bibitem{122} P. Sebbah, O.Legrand, B.A. van Tiggelen, and A.Z. Genack, Phys. Rev. E {\bf 56}, 3619 (1997).
\bibitem{123} P. Sebbah, O. Legrand, and A.Z. Genack, Phys. Rev. E {\bf 59}, 2406 (1999).
\bibitem{124} J.H. Li, {\it Interfacial scattering and microparticle resonances in random media},
Ph.D. thesis, City University of New York, (1993).
\bibitem{125} K. Busch and C. M. Soukoulis, Phys. Rev. B {\bf 54}, 893 (1996).
\bibitem{126} A.Z. Genack, A.A. Chabanov, P. Sebbah, and B.A. van Tiggelen,
in {\it Wave Scattering in Complex Media: from Theory to
Applications}, eds. B.A. van Tiggelen and S. Skipetrov, (Kluwer
Academic, Dordrecht, 2003).
\bibitem{127} C.W.J. Beenakker, in {\it Photonic Crystals and Light Localization in the 21st Century},
ed. C.M. Soukoulis, (Kluwer Acadenic, Dordrecht, 2001).
\bibitem{128} I. Freund, Waves in Random Media {\bf 8}, 119 (1998).
\bibitem{129} A.Z. Genack, P. Sebbah, M. Stoytchev, and B.A. van Tiggelen, Phys. Rev. Lett. {\bf 82}, 715 (1999).
\bibitem{130} B.A. van Tiggelen, P. Sebbah, M. Stoytchev, and A.Z. Genack, Phys. Rev. E {\bf 59}, 7166 (1999).
\bibitem{131} H. Schomerus, K.J.H. van Bemmel, and C.W.J. Beenakker, Phys. Rev. E {\bf 63}, 026605 (2001).
\bibitem{132} A. Lagendijk, J. Gomez-Rivas, A. Imhof, F.J.P. Schuurmans, and R. Sprik, in
{\it Photonic Crystals and Light Localization in the 21st
Century}, ed. C.M. Soukoulis, (Kluwer Academic, Dordrecht, 2001).
\bibitem{133} A.A. Chabanov and A.Z. Genack, Phys. Rev. Lett. {\bf 87}, 233903 (2001).
\bibitem{134} C. Texier and A. Comtet, Phys. Rev. Lett. {\bf 82}, 4220 (1999); 
A. Comtet and C. Texier, J. Phys. A {\bf 30}, 8017 (1997).
\bibitem{135} C.J. Bolton-Heaton, C.J. Lambert, V.I. Falko, V. Prigodin, and A.J. Epstein, Phys. Rev. B {\bf 60}, 10569 (1999).
\bibitem{136} A.Ossipov, T. Kottos, and T. Geisel, Phys. Rev. B {\bf 61}, 11411 (2000).
\bibitem{137} B.L. Altshuler and V.N. Prigodin, JETP Lett. {\bf 47}, 43 (1988).
\bibitem{138} Y.V. Fyodorov, JETP Lett. {\bf 78}, 250 (2003).
\bibitem{139} H. Schomerus, Phys. Rev. E {\bf 64}, 026606 (2001).
\bibitem{140} R.G. Newton, {\it Scattering Theory of Waves and Particles}, (Springer-Verlag, New York, 1982), Sec. 11.13.
\bibitem{141} B.A. van Tiggelen and E. Kogan, Phys. Rev. A {\bf 49}, 708 (1994).
\bibitem{142} E.P. Wigner, Phys. Rev. {\bf 98}, 145 (1955).
\bibitem{143} F.T. Smith, Phys. Rev. {\bf 118}, 349 (1960).
\bibitem{144} Y.V. Fyodorov and H.J. Sommers, J. Math. Phys. {\bf 38}, 1918 (1997).
\bibitem{145} P.W. Brouwer, K.M. Fram, and C.W.J. Beenakker, Phys. Rev. Lett. {\bf 78}, 4737 (1997).
\bibitem{146} C.W.J. Beenakker and P.W. Brouwer, Physica E {\bf 9}, 463 (2001).
\bibitem{147} H.J. Sommers, D.V. Savin, and V.V. Sokolov, Phys. Rev. Lett. {\bf 87}, 094101 (2001).
\bibitem{148} F. Steinbach, A. Ossipov, T. Kottos, and T. Geisel, Phys. Rev. Lett. {\bf 85}, 4426 (2000).
\bibitem{149} T. Kottos and M. Weiss, Phys. Rev. Lett. {\bf 89}, 056401 (2002).
\bibitem{150} A. Ossipov and Y.V. Fyodorov, Phys. Rev. B {\bf 71}, 125133 (2005).
\bibitem{151} J.A. Mendez-Bermudez and T. Kottos, Phys. Rev. B {\bf 72}, 064108 (2005).
\bibitem{152} A.A. Chabanov, B. Hu, and A.Z. Genack, Phys. Rev. Lett. {\bf 93}, 123901 (2004).
\bibitem{153} P. Sebbah, B. Hu, A.Z. Genack, R. Pnini and B. Shapiro, Phys. Rev. Lett. {\bf 88}, 123901 (2002).
\bibitem{154} A.A. Chabanov, N.P. Tregoures, B.A. van Tiggelen, and A.Z. Genack, Phys. Rev. Lett. {\bf 92}, 173901 (2004).
\bibitem{155} J.W. Goodman, {\it Statistical Optics}, (John Wiley, New York, 2000).
\bibitem{156} A.A. Chabanov and A.Z. Genack, cond-mat/0502334 (2005).

\end{thebibliography}
\end{document}